\documentclass[pre,amsmath,amssymb,twocolumn,superscriptaddress]{revtex4}

\usepackage{amsmath,amssymb}
\usepackage[usenames]{color}
\usepackage[dvipsnames]{xcolor}
\usepackage{amssymb}
\usepackage{dsfont}
\usepackage{grffile}
\usepackage[pdftex]{graphicx}
\usepackage{amsmath, amstext, amssymb, amsfonts, amsxtra}
\usepackage{textcomp}
\usepackage{xspace}
\DeclareSymbolFontAlphabet{\amsmathbb}{AMSb}
\usepackage{color}

\newcommand{\tr}{{\rm tr}}

\newcommand{\brasil}{Departamento de F\'{\i}sica--Instituto de Ci\^{e}ncias Exatas, Universidade Federal de Minas Gerais, CP 702, 30.161-970 Belo Horizonte MG, Brazil}

\begin{document}

\title{One-way street for the energy current: A ubiquitous phenomenon in boundary driven quantum spin chains}

\author{Deborah Oliveira}
\author{Emmanuel Pereira}
\email{emmanuel@fisica.ufmg.br}
\affiliation{\brasil}
\author{Humberto C. F. Lemos}
\email{humbertolemos@ufsj.edu.br}
\affiliation{Departamento de Estat\'{\i}stica, F\'{\i}sica e Matem\'atica, CAP --
Universidade Federal de S\~ao Jo\~ao del-Rei, 36.420-000, Ouro Branco, MG, Brazil}

\begin{abstract}
Focusing on the description of nontrivial properties of the energy transport at quantum scale, we investigate asymmetrical quantum spin chains described by boundary-driven
$\mathit{XXZ}$ and $\mathit{XXX}$ Heisenberg models. We search for symmetries properties of the Lindblad master equation related to the dynamics of the system in order to establish properties of the steady state.
Under rather general assumptions for the target polarization at the boundaries, we show the occurrence of an effect related to (but stronger than) energy rectification, namely, the one-way street
phenomenon, which is the existence of an unique way for the energy flow. Precisely, the energy current does not change in magnitude and direction as we invert the baths at the boundaries: its direction is
completely determined by the asymmetry in the bulk of the chain. The results follow independent of the system size and of the transport regime. Our findings show the ubiquitous occurrence of the one-way street
phenomenon for the energy flow in boundary-driven spin systems and, we believe, they shall be an useful contribution to the area devoted to the investigation and building of efficient quantum devices
used to control and manipulate the energy current.

\end{abstract}


\maketitle

{\it Introduction:} Understanding  the properties of the energy transport at quantum scale is a problem of considerable theoretical and experimental interest that is taking increasing attention in recent years.

The emerging field of quantum thermodynamics urges to the detailed theoretical study of the quantum transport properties, in particular, of the quantum energy currents. Moreover, the amazing on-going progress in experimental manipulations of small quantum systems makes mandatory the theoretical investigation of nonequilibrium features of quantum systems, in particular, their transport characteristics, directly related to the understanding of their behavior out of equilibrium.

Some specific problems of theoretical and experimental importance appear in this context, for example, the possibility of building quantum thermal rectifiers, i.e., the possibility of finding systems with a preferential direction for the energy flow. The thermal rectifier, or thermal diode, is a system in which the magnitude of the energy current changes as we invert the device between two baths. Its investigation is motivated by the success of its electronic analog, the electrical diode, which, together with transistor and other related nonlinear solid state devices, were responsible for the amazing development of modern electronics, with impact in our daily lives. In fact, the interest in energy rectification is an old problem: it appears already within the study of simpler classical models describing the heat conduction and many works are devoted to the theme
\cite{LiRMP, Casati+, Casati2, BHu, Prapid, Psuf+, P2}.

In short, we stress, it is clear the general interest in the investigation of the energy transport, importantly in the quantum scale. In particular, a recurrent problem is the search of conditions which allow an asymmetric transport, i.e., a preferential direction for the currents.

In the present work, we investigate the energy current in boundary-driven quantum spin Heisenberg ($\mathit{XXX}$) and $\mathit{XXZ}$ chains, the archetypal models of open quantum systems, and so, a problem with significance to several areas:
nonequilibrium statistical physics, condensed matter, optics, cold atoms, quantum information, etc. \cite{BP}. As usual, we consider the dynamics given by a Lindblad master equation (LME). Under rather general assumptions for the target
polarization at the boundaries, we show the existence of an intriguing one-way street phenomenon for the energy current, i.e., in asymmetric chains (e.g., graded systems) the direction of the energy current is completely determined by the asymmetry in the bulk of the system, precisely, the direction of the energy flow does not change as we invert the baths at the boundaries. It is important to recall that, in these boundary-driven quantum systems, the
energy current involves heat and also power (work), and so, no thermodynamic inconsistency is present here \cite{FBarra, P18, G-NJP}. More details are presented ahead. To establish our results we use symmetries of the density matrix, of the LME, and show the energy current  properties in the steady state. That is, our strategy is inspired by the work of Popkov and Livi \cite{PopLi}. Our results are independent of the system size and of the existence of
anomalous, normal or ballistic transport on the chain.

The existence of this one-way street phenomenon has been already shown for the simpler case of target $\sigma^{z}$ polarization at the edges and another quite specific case \cite{Prap17}. Here, in the present work, we extend
the existence of such phenomenon for rather general cases of boundary spin polarizations, proving that the phenomenon ubiquitously holds in boundary-driven quantum spin chains.

It is important to emphasize that the existence of asymmetry in the bulk of the chain is not a guarantee for the presence of asymmetry in the energy flow. For example, for the case of classical chains of harmonic oscillators with
self-consistent inner stochastic reservoirs, an effective model used to study heat conduction (Fourier's law holds in such model \cite{BLL}, a footprint of anharmonicity, since the law is absent in purely harmonic models \cite{RLL}), it is
rigorously proved that, for any asymmetric chain, there is no thermal rectification, i.e., there is no asymmetry in the magnitude of the heat flow as we invert the baths at the edges \cite{PLA}.

{\it Models and Currrents:}
Now we introduce the spin models. We assume, in what follows, $\hbar=1$. We take quantum 1/2 spin chains and we first consider asymmetric $\mathit{XXZ}$ models with Hamiltonians
\begin{equation}
\mathcal{H} =  \sum_{i=1}^{N-1}\left\{ \alpha\left(\sigma_{i}^{x}\sigma_{i+1}^{x} + \sigma_{i}^{y}\sigma_{i+1}^{y}\right)  + \Delta_{i,i+1}\sigma_{i}^{z}\sigma_{i+1}^{z} \right\}
 ~, \label{hamiltonian}
\end{equation}
where $\sigma_{i}^{\beta}$ ($\beta = x, y, z$) are the Pauli matrices. The anisotropy parameters $\Delta_{i,i+1}$ are assumed to be asymmetrically distributed, for example, with a graded distribution:
$\Delta_{1,2}<\Delta_{2,3}<\cdots<\Delta_{N-1,N}$.

As usual, the dynamics of the system is given by Lindblad master equations (LME) for the density matrix
\begin{equation}
\frac{d\rho}{d t} = i[\rho, \mathcal{H}] + \mathcal{L}(\rho) ~,\label{master}
\end{equation}
where the dissipator $\mathcal{L}(\rho)$ is local here, and describes the coupling with the baths. It is given by
\begin{eqnarray}
\mathcal{L}(\rho) &=& \mathcal{L}_{L}(\rho) + \mathcal{L}_{R}(\rho) ~, \nonumber\\
\mathcal{L}_{L,R}(\rho) &=& \sum_{s=\pm} L_{s}\rho L_{s}^{\dagger} -
\frac{1}{2}\left\{ L_{s}^{\dagger}L_{s} , \rho \right\} ~,\label{dissipator}
\end{eqnarray}
$\{\cdot,\cdot\}$ above describes the anti-commutator. These models are recurrently studied: see, e.g., Ref.\cite{Prosen+} and references therein.
For $\mathcal{L}_{L}$, in the simpler case of a $\mathit{XXZ}$ chain with target $\sigma^z$ polarization at the edges,  analyzed in Ref.\cite{Prap17} and several other works, we have
\begin{equation}
L_{\pm} = \sqrt{\frac{\gamma}{2}(1 \pm f_{L})} \sigma_{1}^{\pm} ~\label{dissipator2},
\end{equation}
where $\sigma_{j}^{\pm}$ are the spin creation and annihilation operators $\sigma_{j}^{\pm} = (\sigma_{j}^{x} \pm i\sigma_{j}^{y})/2$~;
and similarly  for $\mathcal{L}_{R}$, but with $\sigma_{N}^{\pm}$ and $f_{R}$ replacing  $\sigma_{1}^{\pm}$ and $f_{L}$. In the previous expressions,  $\gamma$ denotes the coupling strength to the spin baths; $f_{L}$ and $f_{R}$ describe the
driving strength, and they are related to the polarization of extra spin  at the boundaries:
$$f_{L} = \langle\sigma_{0}^{z}\rangle, ~~~~  f_{R} = \langle\sigma_{N+1}^{z}\rangle .$$
Ahead, we will investigate in details more general and intricate dissipators.

The expressions for the spin and energy currents can be obtained  from the LME for the dynamics and a continuity equation, as precisely described in Ref.\cite{Mendoza-A}. We have, at site $j$, for the magnetization (spin) current
\begin{equation}
\langle J_{j} \rangle = 2\alpha \langle \sigma_{j}^{x} \sigma_{j+1}^{y} - \sigma_{j}^{y}\sigma_{j+1}^{x} \rangle ~.
\end{equation}
If we add in the Hamiltonian (\ref{hamiltonian}) a term giving the interaction with an external magnetic field,
$$\sum_{j=1}^{N}B_{j}\sigma_{j}^{z}~,$$
for the energy current we obtain (again, details in Ref.\cite{Mendoza-A}),
\begin{eqnarray}
\langle F_{j}\rangle &=& \langle F_{j}^{\mathit{XXZ}}\rangle + \langle F_{j}^{B}\rangle ~,\nonumber\\
\langle F_{j}^{\mathit{XXZ}} \rangle &=&  2\alpha \langle \alpha \left( \sigma_{j-1}^{y}\sigma_{j}^{z} \sigma_{j+1}^{x} - \sigma_{j-1}^{x}\sigma_{j}^{z}\sigma_{j+1}^{y}\right) \nonumber\\
&& + \Delta_{j-1,j}\left( \sigma_{j-1}^{z}\sigma_{j}^{x} \sigma_{j+1}^{y} - \sigma_{j-1}^{z}\sigma_{j}^{y}\sigma_{j+1}^{x}\right) \nonumber\\
&& + \Delta_{j,j+1}\left( \sigma_{j-1}^{x}\sigma_{j}^{y} \sigma_{j+1}^{z} - \sigma_{j-1}^{y}\sigma_{j}^{x}\sigma_{j+1}^{z}\right)\rangle  ~,\nonumber\\
\langle F_{j}^{B} \rangle &=& \frac{1}{2} B_{j}\langle J_{j-1} + J_{j}\rangle ~. \label{currents}
\end{eqnarray}
We need to make an important remark now. In most of the studies of $\mathit{XXZ}$ chains, the authors take homogeneous, symmetric chains. In such a case, considering the energy investigation, both  direct computation or symmetry arguments lead to $\langle F^{\mathit{XXZ}}_{j}\rangle = 0$ \cite{Mendoza-A, PopLi}. Consequently, the remaining energy flow is proportional to the magnetization current, and it vanishes in the absence of an external magnetic field $B=0$. But that is not the scenario for asymmetric chains. We give a concrete example. In a previous work \cite{SPL}, the density matrix is analytically computed for a small chain of three spins, and the spin and energy currents are precisely determined for
the case of target $\sigma^{z}$ polarizations, with $f_{L}=f$ and $f_{R}=-f$ at the edges. In particular, in the steady state, for $\langle F \rangle \equiv \langle F_{j}\rangle$, an exact, huge expression is determined for the three sites chain with $\alpha=1$, $\Delta_{1,2} = \Delta - \delta$, $\Delta_{2,3} = \Delta + \delta$, and $B_{j}= B$. For simplicity and clearness, we write below the dominant terms considering an expansion in powers of $f$, the driving strength, and of $\delta$, the asymmetry
parameter. We have
\begin{eqnarray*}
\langle F \rangle &=& Bf \left( \frac{912}{969 + 48\Delta^{2}}\right) \\
 && + f^{2}\delta \left( \frac{32(20224\Delta^{4} + 64256\Delta^{2} -1083)}{(51 + 16\Delta^{2})(323 + 16\Delta^{2})^{2}}\right) ~.
\end{eqnarray*}
From the expression above, we see that the energy current is nonvanishing even for $B=0$, in clear contrast with the behavior present in symmetrical chains, which are usually investigated. Again, for $B=0$, the value (direction and magnitude) of the energy current does not change as we invert the baths, i.e., as we change $f$ by $-f$ (it indeed follows, we stress, for the complete expression beyond $\mathcal{O}(f^{2})$ \cite{SPL} - the complete current is an even function of $f$). That is the one-way street phenomenon, directly computed in Ref.\cite{SPL} and derived by symmetry arguments in Ref.\cite{Prap17}.

Note that, here, we propose to show the general occurrence of such an effect, i.e., the occurrence of an energy current whose direction depends only on the
asymmetry of the chain (it does not invert as we invert the baths at the
boundaries of the chain), and such an effect is, say, stronger than the
energy rectification, even the perfect rectification, that means current in
one direction and absence of current as we invert the baths.

Another important remark is convenient here. In some works with boundary driven quantum systems, the energy current is wrongly taken as the heat current. In such a case, thermodynamic inconsistencies are claimed to occur, such as
a heat flow from the colder to the hotter bath without other interventions \cite{Levy-EPL}. But, in fact, as clearly explained in Ref.\cite{FBarra}, besides heat we also have work in the boundary driven processes, and so, thermodynamic consistency is recovered. See also Ref.\cite{P18} for the related analysis in these $\mathit{XXZ}$ chains, and Ref.\cite{G-NJP} for general considerations (in particular, responding the false inconsistency raised in Ref.\cite{Levy-EPL}).

{\it Results:}
The present work is devoted to extend the one-way street phenomenon for the energy current to $\mathit{XXZ}$ and $\mathit{XXX}$ Heisenberg asymmetric chains with general polarization at the edges. In other words, here, after a considerable algebraic
 work, we show that such ``strange'' phenomenon is ubiquitous in boundary driven asymmetric quantum spin chains.

We proceed by taking the $\mathit{XXZ}$ chain as given by Eqs.(\ref{hamiltonian}, \ref{master}, \ref{dissipator}). We first consider the case in which one edge of the chain is target in a given polarization, say, $\sigma^{x}$, and the other edge involves a polarization with arbitrary twisting angle in the XY plane. Precisely, now we take the dissipators as
\begin{eqnarray}
K^{L}_{\pm} &=& \sqrt{\gamma(1 \pm f)}\left(\frac{\sigma_{1}^{y} \pm i\sigma_{1}^{z}}{2} \right)~, \nonumber\\
K^{R}_{\pm} &=& \sqrt{\gamma(1 \mp f)}\left(\frac{\cos(\theta)\sigma_{N}^{x} +\sin(\theta)\sigma_{N}^{y} \pm i\sigma_{N}^{z}}{2}\right) .
\end{eqnarray}
The Heisenberg ($\mathit{XXX}$) version of such model is investigated in Refs.\cite{PPS-PRE, PKS-PRE, LandiK}. Our strategy here is to exploit the symmetries in the steady state of the LME in order to show that, if $\rho$ is a steady state
solution, then there exists a unitary transformation $U$ such that $U\rho U^{\dagger}$ is a solution of the steady state LME with inverted baths. Moreover, for the LME with inverted baths, the energy current in the absence of
external magnetic field $B$ is the same. In resume, we want to show for the heat current
\begin{eqnarray}
\langle F \rangle &=& \langle F^{\mathit{XXZ}} \rangle \equiv \tr\left( \rho  F_{j}^{\mathit{XXZ}}\right) = \tr \left(\rho U^{-1}F_{j}^{\mathit{XXZ}}U\right) \nonumber\\
 &=& \tr\left( U\rho U^{-1}F_{j}^{\mathit{XXZ}}\right) = \langle F^{\mathit{XXZ}} \rangle_{\rm inv. baths} ~.
 \end{eqnarray}

We make an important statement here. By using the approach described in
Ref.\cite{EvansProsen}, we can prove the uniqueness for the steady state of
all the LME treated in this present work. Thus, if $\rho$ is the steady
distribution of the initial system, then $U\rho U^{-1}$ is the unique steady
distribution of the system with inverted baths. Another important remark: for the studies of symmetries in the LME, our operator $U$ is indeed the tensorial product over all $N$ sites of
$2\times 2$ unitary transformations (they are the same transformation, but each one acts on one site)
\begin{equation}
U = u\otimes u\otimes \ldots \otimes u ~.
\end{equation}

For more details about the desired transformation $U$, we note that, in the steady state the LME reads
$$
0 = -i[\mathcal{H}, \rho] + \mathcal{L}(\rho) ~.
$$
Hence, we must find $U$ such that
\begin{equation}
\mathcal{H} = U\mathcal{H}U^{\dagger}, ~~~~ \mathcal{L}_{\rm inv. baths}\left(U\rho U^{\dagger}\right) = U \mathcal{L} U^{\dagger} ~.
\end{equation}
Similarly results with the change $U \leftrightarrow U^{\dagger}$ (e.g., $\mathcal{H} = U^{\dagger}\mathcal{H}U$). I.e., for this first case, we need to present $U$ unitary such that (discarding the change $1\leftrightarrow N$)
$$
U K^{L,R}_{\pm} U^{\dagger} = K^{R,L}_{\mp} ~.
$$
To find this desired $U$, or each $u$, we start from a general representation for an unitary matrix $u$
\begin{eqnarray*}
u \equiv  \left(\begin{array}{cc} a & b \\
 -e^{i\varphi}b^{*} & e^{i\varphi}a^{*} \end{array} \right)
 ~,
 \end{eqnarray*}
where $a$ and $b$ are complex numbers; $a^{*}$ is the complex conjugated; $\varphi$ is real; and $|a|^{2} + |b|^{2} =1$. Then, we investigate if it is possible to describe $U$ which inverts the baths and that satisfies all the
previous relations described above. Of course, we do not present the algebraic manipulations carried out to find the desired $U$. But the reader can check {\it a posteriori} that the desired relations follow with the presented $U$.

For the first case in which the dissipators are given by $K^{R,L}_{\pm}$, i.e., for the case of one edge of the chain with a $\sigma^{x}$ polarization, and the other edge with a polarization in a arbitrary direction in the XY
plane, after a tedious algebraic work we find $u$ as given by
\begin{equation}
u^{(1)} = \frac{1}{\sqrt{2}}\begin{pmatrix}
    0 & 1+i\\
    -e^{i\theta}(1-i) & 0
\end{pmatrix} ~~.
\end{equation}
For emphasis, we repeat that, with such a matrix, it follows $U^{(1)}F^{\mathit{XXZ}} U^{(1)\dagger} = F^{\mathit{XXZ}}$, that is, the energy current is the same as we invert the baths. As a further observation, we note that
\begin{equation*}
    u^{(1)}\sigma^{x}u^{(1)\dagger} = \begin{pmatrix}
     0 & -\sin{\theta} -i\cos{\theta}\\
     -\sin{\theta} + i\cos{\theta}  & 0
     \end{pmatrix}~~,
\end{equation*}
\begin{equation*}
    u^{(1)}\sigma^{y}u^{(1)\dagger} = \begin{pmatrix}
     0 & \cos{\theta} - i\sin{\theta}\\
     \cos{\theta}+i\sin{\theta} & 0
     \end{pmatrix}~~,
\end{equation*}
\begin{equation*}
    u^{(1)}\sigma^{z}u^{(1)\dagger} = -\sigma^{z} ~~.
\end{equation*}
Although we have awkward spin transformations, it still follows that $U^{(1)}\mathcal{H}U^{(1)\dagger} = \mathcal{H}$ as well.

In the next step, we consider a general case of LME involving several dissipators: $L_{1}, L_{2}, V_{1}, V_{2}, W_{1}, W_{2}$ acting on the first site, and $L_{3}, L_{4}, V_{3}, V_{4}, W_{3}, W_{4}$ acting on
the site $N$. Precisely,
\begin{eqnarray}
L_{1} &=& \alpha( \sigma_{1}^{x} + i\sigma_{1}^{y} )~, ~~~~L_{3} = \beta( \sigma_{N}^{x} + i\sigma_{N}^{y} ) ~, \nonumber \\
L_{2} &=& \beta( \sigma_{1}^{x} - i\sigma_{1}^{y} )~, ~~~~L_{4} = \alpha( \sigma_{N}^{x} - i\sigma_{N}^{y} ) ~, \nonumber \\
V_{1} &=& p( \sigma_{1}^{y} + i\sigma_{1}^{z} )~, ~~~~V_{3} = v( \sigma_{N}^{y} + i\sigma_{N}^{z} ) ~, \nonumber \\
V_{2} &=& q( \sigma_{1}^{y} - i\sigma_{1}^{z} )~, ~~~~V_{4} = u( \sigma_{N}^{y} - i\sigma_{N}^{z} ) ~, \nonumber \\
W_{1} &=& u( \sigma_{1}^{z} + i\sigma_{1}^{x} )~, ~~~~W_{3} = q( \sigma_{N}^{z} + i\sigma_{N}^{x} ) ~, \nonumber \\
W_{2} &=& v( \sigma_{1}^{z} - i\sigma_{1}^{x} )~, ~~~~W_{4} = p( \sigma_{N}^{z} - i\sigma_{N}^{x} ) ~.
\end{eqnarray}
The parameters $\alpha, \beta$, $p, q, u, v$ above can be taken as nonegative
real numbers. Note that we have the same parameters acting both on first and last sites of the spin chain, but they are linked to different target polarization operators at each boundary.
The operators $L_{k}, V_{k}$, and $W_{k}$, when taken alone, target polarization along the axes $z, x$ and $y$, respectively.

Again, the procedure is the same, and we find $U^{(2)}$ such that $U^{(2)}\rho U^{(2)\dagger}$ satisfies the LME with inverted baths and $U^{(2)}F^{\mathit{XXZ}}U^{(2)\dagger} = F^{\mathit{XXZ}}$, i.e., the one-way street phenomenon holds. We obtain
\begin{equation}
    u^{(2)} =\frac{1}{\sqrt{2}}\begin{pmatrix}
    0 & -1+i\\
    1+i & 0
    \end{pmatrix} ~~.
\end{equation}
Moreover, we have
\begin{eqnarray*}
   u^{(2)}\sigma^{x}u^{(2)\dagger} &=& -\sigma^{y}~, ~~~~   u^{(2)}\sigma^{y}u^{(2)\dagger} = -\sigma^{x}~,\\
    u^{(2)}\sigma^{z}u^{(2)\dagger} &=& -\sigma^{z}~.
\end{eqnarray*}

We turn, now, to the investigation of asymmetric Heisenberg $\mathit{XXX}$ models, i.e., we extend the asymmetry distribution also to the $x$ and $y$ coordinates. Precisely, we consider the Hamiltonian
\begin{equation}
\mathcal{H} =  \sum_{i=1}^{N-1} \alpha_{i}\left(\sigma_{i}^{x}\sigma_{i+1}^{x} + \sigma_{i}^{y}\sigma_{i+1}^{y}  + \sigma_{i}^{z}\sigma_{i+1}^{z} \right) ~, \label{hamiltonian2}
\end{equation}
where $\alpha_{i}$ is assumed to be asymmetrically distributed.

In this case, we need to rewrite the expressions for the currents. For the spin flow we have
\begin{equation}
\langle J_{j} \rangle = 2\alpha_{j} \langle \sigma_{j}^{x} \sigma_{j+1}^{y} - \sigma_{j}^{y}\sigma_{j+1}^{x} \rangle ~.
\end{equation}
And, for the energy current, we obtain
\begin{eqnarray}
\langle F_{j}^{\mathit{XXZ}} \rangle &=&  2\alpha_{i-1} \alpha_{i}\langle \left( \sigma_{j-1}^{y}\sigma_{j}^{z} \sigma_{j+1}^{x} - \sigma_{j-1}^{x}\sigma_{j}^{z}\sigma_{j+1}^{y}\right) \nonumber\\
&& + \left( \sigma_{j-1}^{z}\sigma_{j}^{x} \sigma_{j+1}^{y} - \sigma_{j-1}^{z}\sigma_{j}^{y}\sigma_{j+1}^{x}\right) \nonumber\\
&& + \left( \sigma_{j-1}^{x}\sigma_{j}^{y} \sigma_{j+1}^{z} - \sigma_{j-1}^{y}\sigma_{j}^{x}\sigma_{j+1}^{z}\right)\rangle  ~.\nonumber\\
\label{currents2}
\end{eqnarray}

In relation to the dissipators of the LME, we start again with one of the edges of the chain target in a given polarization, say $\sigma^{z}$, and the other edge with an arbitrary polarization in the plane ZX.
\begin{eqnarray}
D^{L}_{\pm} &=& \sqrt{\gamma(1 \pm f)}\left(\frac{\sigma_{1}^{x} \pm i\sigma_{1}^{y}}{2} \right)~, \nonumber\\
D^{R}_{\pm} &=& \sqrt{\gamma(1 \mp f)}\left(\frac{\cos(\theta)\sigma_{N}^{x} +\sin(\theta)\sigma_{N}^{z} \pm i\sigma_{N}^{y}}{2}\right) .
\end{eqnarray}

Again, after considerable algebraic manipulations, we find the transformation
$U^{(3)}$ that inverts the baths and shows the one-way street phenomenon for the
energy current. We have
\begin{equation}
u^{(3)}=\frac{i}{\sqrt{2}}\begin{pmatrix}
\sqrt{1-\cos{\theta}} & \sqrt{1+\cos{\theta}} \\
\sqrt{1+\cos{\theta}} & -\sqrt{1-\cos{\theta}}
\end{pmatrix} ~.
\end{equation}

It is interesting to note the intricate transformation for the spin variables
here,
\begin{equation*}
u^{(3)}\sigma^{x}u^{(3)\dagger} = \begin{pmatrix}
\sin{\theta} & \cos{\theta}\\
\cos{\theta} & -\sin{\theta}
\end{pmatrix}~,
\end{equation*}
\begin{equation*}
u^{(3)}\sigma^{y}u^{(3)\dagger} = -\sigma^{y}~,
\end{equation*}
\begin{equation*}
u^{(3)}\sigma^{z}u^{(3)\dagger} = \begin{pmatrix}
-\cos{\theta} & \sin{\theta}\\
\sin{\theta} & \cos{\theta}
\end{pmatrix}~.
\end{equation*}
Anyway, as said, we still have $U^{(3)}\mathcal{H}U^{(3)\dagger} = \mathcal{H}$, and $U^{(3)}F U^{(3)\dagger} = F$.

Finally, we consider the case of several dissipators, which, alone, target polarization along the axes $x, y$ and $z$. We take
\begin{eqnarray}
L_{1} &=& \alpha( \sigma_{1}^{x} + i\sigma_{1}^{y} )~, ~~~~L_{3} = v( \sigma_{N}^{x} + i\sigma_{N}^{y} ) ~, \nonumber \\
L_{2} &=& \beta( \sigma_{1}^{x} - i\sigma_{1}^{y} )~, ~~~~L_{4} = u( \sigma_{N}^{x} - i\sigma_{N}^{y} ) ~, \nonumber \\
V_{1} &=& p( \sigma_{1}^{y} + i\sigma_{1}^{z} )~, ~~~~V_{3} = q( \sigma_{N}^{y} + i\sigma_{N}^{z} ) ~, \nonumber \\
V_{2} &=& q( \sigma_{1}^{y} - i\sigma_{1}^{z} )~, ~~~~V_{4} = p( \sigma_{N}^{y} - i\sigma_{N}^{z} ) ~, \nonumber \\
W_{1} &=& u( \sigma_{1}^{z} + i\sigma_{1}^{x} )~, ~~~~W_{3} = \beta( \sigma_{N}^{z} + i\sigma_{N}^{x} ) ~, \nonumber \\
W_{2} &=& v( \sigma_{1}^{z} - i\sigma_{1}^{x} )~, ~~~~W_{4} = \alpha( \sigma_{N}^{z} - i\sigma_{N}^{x} ) ~,
\end{eqnarray}
(note that they involve a combination different from the previous one in the $\mathit{XXZ}$ case). Again, the  parameters $\alpha, \beta$, $p, q, u, v$ above can be taken as nonegative
real numbers.

The desired matrix changing the baths is now found as
\begin{equation}
u^{(4)} =\frac{1}{\sqrt{2}}\begin{pmatrix}
i & -1\\
1 & -i
\end{pmatrix} ~.
\end{equation}

For the spin transformations we get
\begin{eqnarray*}
u^{(4)}\sigma^{x}u^{(4)\dagger}  &=& - \sigma^{x}
 ~,~~~~~~u^{(4)}\sigma^{y}u^{(4)\dagger} = - \sigma^{z} \\
u^{(4)}\sigma^{z}u^{(4)\dagger} &=& - \sigma^{y}~.
\end{eqnarray*}

Once more, we have $U^{(4)}\mathcal{H}U^{(4)\dagger} = \mathcal{H}$, and $U^{(4)}F U^{(4)\dagger} = F$.

{\it Conclusion:}
It is worth to recall that asymmetric systems, as considered in this work, are
not only theoretical proposals. For example, there is a proliferation of
graded materials in nature, i.e., inhomogeneous systems whose structure
changes gradually in space, and they can be also manufactured. There is a
great interest for such materials in many areas: optics, mechanical
engeneering, material science, etc. \cite{graded}. Moreover, a simple example
of graded thermal rectifier has been already built: a carbon and boron
nitride nanotube, inhomogeneously coated with heavy molecules \cite{Chang}.

We believe that we shall see experimental realizations of such asymmetrical
$\mathit{XXZ}$ and Heisenberg ($\mathit{XXX}$) models soon. We recall that it has been already
shown the possibility to engineer $\mathit{XXZ}$ quantum spin Hamiltonians with different
values for the inner parameters $\alpha$ and $\Delta$ \cite{Endres, Barredo}. Moreover,
Heisenberg ($\mathit{XXX}$) and $\mathit{XXZ}$ models appear related to recent experimental works
with Rydberg atoms in optical traps \cite{Duan, Nguyen}.

We show here that an interesting effect related to (but stronger than) rectification is of ubiquitous occurrence in boundary driven quantum spin systems with target polarization,
the archetypal models of quantum spin nonequilibrium physics. In order to stress such a generalization of our results, a final technical comment is pertinent.  Here, we show the one-way street phenomenon for a $\mathit{XXZ}$ chain in which one edge of the chain is target in a given polarization, say, $\mathit{x}$ direction, and the other edge involves a polarization with arbitrary twisting angle (described as $\theta$ and taking any value) in the $\mathit{XY}$ plane, for example. We also show the phenomenon for arbitrary choices of six operators, i.e.,  pairs of three kind of operators that, when taken alone,  target polarization along the axes $\mathit{x}$, $\mathit{y}$ and $\mathit{z}$ (one pair targets along $\mathit{x}$, the other $\mathit{y}$, etc.).  That is why we repeatedly say that we are considering general polarizations. Our findings are not resumed as new examples of specific polarization involving quite specific angles or some specific axis. Similar results have been shown for the Heisenberg system.

We are confident that such results will stimulate more theoretical and experimental research on the theme of quantum transport.

\vspace*{1 cm} {\bf Acknowledgments:} This work was partially supported by CNPq (Brazil).




\begin{thebibliography}{32}

\bibitem{LiRMP} N. Li, J. Ren, L. Wang, G. Zhang, P. H\"{a}nggi, and B. Li, Rev.
Mod. Phys. {\bf84}, 1045 (2012).

\bibitem{Casati+} M. Terraneo, M. Peyrard, and G. Casati, Phys. Rev. Lett. {\bf 88}, 094302 (2002).

 \bibitem{Casati2} B. Li, L. Wang, and G. Casati, Phys. Rev. Lett. {\bf 93}, 184301 (2004).

 \bibitem{BHu} B. Hu, L. Yang, and Y. Zhang, Phys. Rev. Lett. {\bf 97}, 124302 (2006).

\bibitem{Prapid} E. Pereira, Phys. Rev. E {\bf 82}, 040101(R) (2010).




\bibitem{Psuf+}  E. Pereira, Phys. Rev. E {\bf 83}, 031106 (2011).

\bibitem{P2} J. Wang, E. Pereira, and G. Casati, Phys. Rev. E  {\bf 86}, 010101 (R) (2012).




\bibitem{BP} H. P. Breuer and F. Petruccione, ``The Theory of Open Quantum Systems'' (Oxford University Press, Oxford, 2002).



\bibitem{FBarra} F. Barra, Sci. Rep. {\bf 5}, 14873 (2015).

\bibitem{P18} E. Pereira, Phys. Rev. E \textbf{97}, 022115 (2018).

\bibitem{G-NJP} G. De Chiara, G. Landi, A. Hewgill, B. Reid, A. Ferraro, A. J. Roncaglia, and M. Antezza, New J. Phys. \textbf{20}, 113024 (2018).


\bibitem{PopLi} V. Popkov and R. Livi, New J. Phys. \textbf{15}, 023030 (2013).

\bibitem{Prap17} E. Pereira, Phys. Rev. E \textbf{95}, 030104 (R) (2017).

\bibitem{BLL} F. Bonetto, J. L. Lebowitz, and J. Lukkarinen, J. Stat. Phys. \textbf{116}, 783 (2004).

\bibitem{RLL} Z. Rieder, J. L. Lebowitz, and E. Lieb, J. Math. Phys. {\bf 8}, 1073 (1967).

\bibitem{PLA} E. Pereira, H. C. F. Lemos, and R. R. \'Avila, Phys. Rev. E \textbf{84}, 061135 (2011).

\bibitem{Prosen+} T. Prosen, Phys. Rev. Lett. {\bf 106}, 217206 (2011); M. \v{Z}nidari\v{c},  Phys. Rev. Lett. {\bf 106}, 220601 (2011); D. Karevski {\textit et al.}, Phys. Rev. Lett. {\bf 110}, 047201 (2013).

\bibitem{Mendoza-A} J. J. Mendoza-Arenas, S. Al-Assam, S. R. Clark, and D. Jaksch, J. Stat. Mech. {\bf 2013}, P07007 (2013).

\bibitem{SPL} L. Schuab, E. Pereira, and G. T. Landi, Phys. Rev. E {\bf 94}, 042122 (2016).

\bibitem{Levy-EPL} A. Levy, and R. Koloff, EPL (Europhysics Letters) \textbf{107}, 20004 (2014).



\bibitem{PPS-PRE} V. Popkov, M. Salerno, and G. M. Schutz, Phys. Rev. E \textbf{85}, 031137 (2012).

\bibitem{PKS-PRE} V. Popkov, D. Karevski, and G. M. Schutz, Phys. Rev. E \textbf{88}, 062118 (2013).

\bibitem{LandiK} G. T. Landi, and D. Karevski, Phys. Rev. B \textbf{91}, 174422 (2015).

\bibitem{EvansProsen} D. Evans, Commun. Math. Phys. {\bf 54}, 293 (1977); T. Prosen, Physica Scripta {\bf 86}, 058511 (2012).

\bibitem{graded} J. P. Huang, and K. W. Yu, Phys Rep. \textbf{431}, 87 (2006).



\bibitem{Chang} C. W. Chang, D. Okawa, A. Majumdar, and A. Zettl, Science \textbf{314}, 1121 (2006).


\bibitem{Endres} M. Endres, H. Bernien, A. Keesling, H. Levine, E. R. Anschuetz, A. Krajenbrink, C. Senko, V. Vuletic, M. Greiner, G. Markus and M. D. Lukin, Science \textbf{354}, 1024 (2016).



\bibitem{Barredo} D. Barredo, S. De L{\'e}s{\'e}leuc, V. Lienhard, T. Lahaye, and A. Browaeys,  Science \textbf{354}, 1021 (2016).

\bibitem{Duan} L.-M. Duan, E. Demler, and M. D. Lukin, Phys. Rev. Lett. \textbf{91}, 090402 (2003).


\bibitem{Nguyen} T. L. Nguyen, J. M.  Raimond, C. Sayrin, R. Corti\~nas, T. Cantat-Moltrecht, F. Assemat, I. Dotsenko, S. Gleyzes, S.  Haroche, G. Roux, Th. Jolicoeur,  and M. Brune, Phys. Rev. X \textbf{8},
 011032 (2018).

























































































































































\end{thebibliography}
\end{document}